# Some New Unifications in Supersymmetry and Higher Dimensional Complex Space

Yi-Fang Chang

*Department of Physics, Yunnan University, Kunming 650091, China*

(e-mail: yifangchang1030@hotmail.com)

**Abstract**

Some new representations of the supersymmetric transformations are derived, and the supermultiplets are introduced. Based on these representations, various formulations (equations, commutation relations, propagators, Jacobi identities, etc.) of bosons and fermions may be unified. On the one hand, the mathematical characteristic of particles is proposed: bosons correspond to real number, and fermions correspond to imaginary number, respectively. Such fermions of even (or odd) number form bosons (or fermions), which is just consistent with a relation between imaginary and real number. The imaginary number is only included in the equations, forms, and matrixes of fermions. It is connected with relativity. On the other hand, the unified forms of supersymmetry are also connected with the statistics unifying Bose-Einstein and Fermi-Dirac statistics, and with the possible violation of Pauli exclusion principle; and a unified partition function is obtained. Therefore, one of the possible developments is the higher dimensional complex space.

## 1. Supersymmetry and its new representations of transformation

Since a basic symmetry between bosons and fermions begins to attach importance in the early days of 1970s, the supersymmetry theory is continuously one of the most active regions in particle physics. The supersymmetry is connected with superspace, superfield and super-transformation, etc. It arouses the supergravity in the gravitational theory. Combining a string model, a well-known superstring is derived. S.Weinberg introduced systematically supersymmetry in his book <The Quantum Theory of Field> Vo.III [1]. Recently, some directions in supersymmetry are discussed: Yang-Mills gauge theory [2-9], the supersymmetry breaking [7,10-19], dark matter and higher dimensional supersymmetric space [6,20-25], the minimal supersymmetric standard model [26-28]. Moreover, supersymmetry and R parity violation and CP asymmetry, and closed-string tachyon condensation, and physics of crypto-supersymmetry field theories are researched [29-31].

In this paper, some new representations of the supersymmetric transformation are derived. Based on these representations, we discuss the super-unification of bosons and fermions, and its mathematical characteristic, physical meaning and a possibly developed direction.

The superfield, for example, a scalar superfield $\Phi(x,\theta)$, includes bose fields $\varphi$ (they may be various scalar fields, vector fields, tensor fields, etc.) and fermi spinor fields $\psi$. The super-gauge transformation between two fields may be represented as [32,33]:

$$\delta\varphi = A\bar{\varepsilon}\psi,$$
$$\delta\psi = -B\varphi\varepsilon, \qquad (1)$$

where the parameter $\bar{\varepsilon}$ is an anticommuting Majorana spinor.

We propose a supposition: A particle supermultiplet is $\Phi = \begin{pmatrix} \varphi \\ ib\bar{\varepsilon}\psi \end{pmatrix}$, then Eq.(1) becomes:

$$\delta\varphi = -iAb^{-1}(ib\bar{\varepsilon}\psi),$$
$$\delta(ib\bar{\varepsilon}\psi) = -ib\bar{\varepsilon}B\varphi\varepsilon. \qquad (2)$$

$$\therefore \begin{pmatrix} \varphi' \\ ib\bar{\varepsilon}\psi' \end{pmatrix} = \begin{pmatrix} 1 & -iAb^{-1} \\ -ib\bar{\varepsilon}B\varepsilon & 1 \end{pmatrix} \begin{pmatrix} \varphi \\ ib\bar{\varepsilon}\psi \end{pmatrix}, \qquad (3)$$

where $\varepsilon$ commute with bose fields $\varphi$, and anticommute with fermi fields $\psi$ [32,33]. Let

$$\alpha_{\mu\nu} = \begin{pmatrix} 1 & -iAb^{-1} \\ -ib\bar{\varepsilon}B\varepsilon & 1 \end{pmatrix} = \begin{pmatrix} 1 & 0 \\ 0 & 1 \end{pmatrix} - i\begin{pmatrix} 0 & Ab^{-1} \\ b\bar{\varepsilon}B\varepsilon & 0 \end{pmatrix}, \quad (4)$$

and $\quad N = C\alpha_{\mu\nu}^{2} = C[\begin{pmatrix} 1 & 0 \\ 0 & 1 \end{pmatrix}^{2} - \begin{pmatrix} A\bar{\varepsilon}B\varepsilon & 0 \\ 0 & b\bar{\varepsilon}B\varepsilon Ab^{-1} \end{pmatrix} - 2i\begin{pmatrix} 0 & Ab^{-1} \\ b\bar{\varepsilon}B\varepsilon & 0 \end{pmatrix}]. \quad (5)$

Of course, we may symmetrically suppose that an equivalent a multiplet of super particles is $\begin{pmatrix} ia\varphi\varepsilon \\ \psi \end{pmatrix}$, or other forms. The multiplet of super particles $\Phi = \begin{pmatrix} \varphi \\ ib\bar{\varepsilon}\psi \end{pmatrix}$ is analogous to a pair of particles of spin s＝1/2, and one of isospin I＝1/2. All of they are quantized doublet. Moreover, it can be extended to a multiplet, e.g.,

$$\begin{pmatrix} \varphi & s=0 \\ ib\bar{\varepsilon}\psi & s=1/2 \\ cA\mu & s=1 \\ id\bar{\varepsilon}\chi & s=3/2 \\ . & . \\ . & . \end{pmatrix}. \quad (6)$$

A transformation among fields of n types should correspond to n × n matrix.

## 2. Unified forms of superfield

Based on these new representations and the supposition, Graded Lie Algebras of bosons and fermions [34] can be unified. GLA are extensions of usual Lie Algebras where a distinction between even and odd elements is introduced: Even elements belong to an ordinary Lie Algebra, and obey usual commutation relations; odd elements obey anticommutation relations among themselves and commutation relations with the even elements of the Lie Algebra [34]. For even elements $A_m$ of a D-dimensional Lie Algebra and odd elements $Q_\alpha$ of a d-dimensional GLA, let

$$\Phi_m = \begin{pmatrix} A_m \\ i\bar{\varepsilon}^m Q_m \end{pmatrix}, \quad (7)$$

then the commutation rules for the (D+d) dimensional GLA are [34]:

$$[A_m, A_n] = f_{mn}^l A_l, \quad (8)$$
$$[A_m, Q_\alpha] = S_{m\alpha}^\beta Q_\beta, \quad (9)$$
$$\{Q_\alpha, Q_\beta\} = F_{\alpha\beta}^m A_m. \quad (10)$$

They can be unified to

$$\begin{pmatrix} A_m \\ i\bar{\varepsilon}^m Q_m \end{pmatrix}(A_n, i\bar{\varepsilon}^n Q_n) - [\begin{pmatrix} A_n \\ i\bar{\varepsilon}^n Q_n \end{pmatrix}(A_m, i\bar{\varepsilon}^m Q_m)]' =$$

$$\begin{pmatrix} A_m A_n - A_n A_m & iA_m \bar{\varepsilon}^n Q_n - i\bar{\varepsilon}^n Q_n A_m \\ i\bar{\varepsilon}^m Q_m A_n - iA_n \bar{\varepsilon}^m Q_m & -\bar{\varepsilon}^m Q_m \bar{\varepsilon}^n Q_n + \bar{\varepsilon}^n Q_n \bar{\varepsilon}^m Q_m \end{pmatrix} = \begin{pmatrix} f_{mn}^l A_l & i\bar{\varepsilon}^n S_{mn}^l Q_l \\ -i\bar{\varepsilon}^m S_{nm}^l Q_l & \bar{\varepsilon}^m \bar{\varepsilon}^n F_{mn}^l A_l \end{pmatrix}, (11)$$

where $\bar{\varepsilon}$ anticommute with Q, and anticommute each other. A right of Eq.(11) may become

$$\begin{pmatrix} f_{mn}^l & 0 \\ 0 & \bar{\varepsilon}^m\bar{\varepsilon}^n F_{mn}^l \end{pmatrix} A_l + i \begin{pmatrix} 0 & \bar{\varepsilon}^n S_{mn}^l \\ -\bar{\varepsilon}^m S_{nm}^l & 0 \end{pmatrix} Q_l . \quad (12)$$

These Jacobi identities can be unified to

$$[\Phi_m,[\Phi_n,\Phi_l]]+[\Phi_l,[\Phi_m,\Phi_n]]+[\Phi_n,[\Phi_l,\Phi_m]]=0, \quad (13)$$

where $\Phi$ may be A or $i\bar{\varepsilon}Q$. Such Eq.(13) includes three known Jocobi identities [34]:

$$[A_m,[A_n,Q_\alpha]]+[Q_\alpha,[A_m,A_n]]+[A_n,[Q_\alpha,A_m]]=0, \quad (14)$$

If $\Phi_m = A_m, \Phi_n = i\bar{\varepsilon}Q_\alpha, \Phi_l = i\bar{\varepsilon}Q_\beta$, Eq.(13) spreads out, rearranges and obtains

$$[A_m,\{Q_\alpha,Q_\beta\}]+\{[Q_\alpha,A_m],Q_\beta\}+\{[Q_\beta,A_m],Q_\alpha\}=0. \quad (15)$$

The super-Jacobi identity

$$[Q_\alpha,\{Q_\beta,Q_\gamma\}]+[Q_\gamma,\{Q_\alpha,Q_\beta\}]+[Q_\beta,\{Q_\gamma,Q_\alpha\}]=0, \quad (16)$$

may be obtained [1]. A commutation relation of super particles may be unified to

$$[\Phi(x),\Phi(x')]=0, \quad (17)$$

which is commutation or anticommutation relation for bosons or fermions, respectively.

Free fermions are described by differential equations of first order:

$$(\gamma_\mu\partial_\mu+m)\psi=0. \quad (18)$$

Free bosons are described by differential equations of second order:

$$(\Box-m^2)\varphi=0. \quad (19)$$

If the supposition still holds, and assume

$$b\bar{\varepsilon}=-i(\gamma_\mu\partial_\mu-m)^{-1}=-(\gamma_\mu p_\mu+im)^{-1}, \quad (20)$$

the two types of equations will be unified to

$$(\Box-m^2)\Phi=0. \quad (21)$$

The propagator of the spinor field is

$$S_F^0(x)=(m-\gamma_\mu\partial_\mu)\Delta_F^0(x), \quad (22)$$

where $\Delta_F^0$ is propagator of boson. The propagator of multiplet $\Phi$ of super particles may be unified to

$$\frac{1}{(2\pi)^4}\frac{i}{(k^2+\mu^2)}=i(b\bar{\varepsilon})^{-1}\Delta_F^0. \quad (23)$$

The different representations between bosons and fermions are equations in quantum mechanics, and are commutation relations and Feynman rules in quantum field theory. In these conditions they all are unified.

Based on the supposition and above discussions, all forms of representations on bosons and fermions in quantum mechanics and quantum field theory seem to be unified, at least for free particles. Probably, they should be extended for the interacting fields.

## 3. Supersymmetry and complex number

In special relativity the Lorentz transformation may be represented as

$$\begin{pmatrix} x' \\ ict' \end{pmatrix}=\gamma\begin{pmatrix} 1 & -iv/c \\ iv/c & 1 \end{pmatrix}\begin{pmatrix} x \\ ict \end{pmatrix}=\gamma\left[\begin{pmatrix} 1 & 0 \\ 0 & 1 \end{pmatrix}+i\frac{v}{c}\begin{pmatrix} 0 & -1 \\ 1 & 0 \end{pmatrix}\right]\begin{pmatrix} x \\ ict \end{pmatrix}. \quad (24)$$

$I=\begin{pmatrix} 1 & 0 \\ 0 & 1 \end{pmatrix}$ is identical matrix, $i\begin{pmatrix} 0 & -1 \\ 1 & 0 \end{pmatrix}=\sigma_2. I^2=1, \sigma^2=1$. So $\left(\frac{\sigma_2}{i}\right)^2=\begin{pmatrix} 0 & -1 \\ 1 & 0 \end{pmatrix}^2=-1$

is analogous to complex number. Let $\bar{\sigma}_2 = \begin{pmatrix} 0 & -1 \\ 1 & 0 \end{pmatrix}, \bar{\sigma}_1 = \sigma_1, \bar{\sigma}_3 = \sigma_3$, i.e., the Pauli matrices divided by imaginary unit i are $\bar{\sigma}$. Similarly, we may define $\bar{\gamma}_\mu$ from Dirac matrices, and matrices $\bar{\gamma}_{1,2,3}^2 = \gamma_{1,2,3}^2 = 1$, and $\bar{\gamma}_{4,5}^2 = -1$ for Majorana representations [35]. These matrices $\bar{\sigma}_i, \bar{\gamma}_\mu$ consist of 1 and −1. Therefore, the Lorentz transformation and supersymmetrical interchange between bosons and fermions can apply matrices.

We propose a mathematical physical law: Bosons correspond to real number, and fermions correspond to imaginary number.

It corresponds to a signature: r= ±I (fermions) or = ±1(bosons). Let r= $\exp(-i\pi\alpha)$, then

$$\alpha = \begin{cases} \pm 1/2 & (r = \mp i) \\ 0 & (r = 1) \\ 1 & (r = -1) \end{cases} \quad (25)$$

Such bosons and fermions consist of even and odd fermions, respectively, which just corresponds to even and odd imaginary numbers are real and imaginary number.

Inference: $\begin{pmatrix} x \\ ict \end{pmatrix}$ corresponds to $\begin{pmatrix} \varphi \\ ib\bar{\varepsilon}\psi \end{pmatrix}$.

Based on the two basic principles of the special relativity, the v<c Lorentz transformation (LT) and $\bar{v}$ >c general Lorentz transformation (GLT) should be derived simultaneously by the classification of the timelike and spacelike intervals. In deriving LT, an additional independent hypothesis has been applied, thus the values of velocity are restricted absolutely. Otherwise, above two symmetrical structures are derived necessarily. Therefore, the present formulations of the special relativity are imperfect [36-38]. From this some fundamental properties between superluminal tachyon and subluminal particle, and spacelike and timelike intervals change each other. For example, those signs of v<c, p<E/c, A<$\varphi$ are just opposite. It corresponds to interchange of some properties between bosons and fermions.

The fermions equations are Dirac equation of four-component (18), which correspond to

$$p_i\psi = -i\hbar\frac{\partial\psi}{\partial x_i}, \quad (26)$$

or Weyl neutrino equations of two-component. They and corresponding Dirac matrices $\gamma_\mu$ and Pauli matrices $\sigma_\alpha$ are linked to imaginary unit i. The bosons equations are Klein-Gordon equation (19), which corresponds to

$$p_i^2\varphi = -\hbar^2\frac{\partial^2\varphi}{\partial x_i^2}, \quad (27)$$

or Proca equation. They are independent of i. Further, the developed Kemmer equations [35], whose forms are analogous to Dirac equations

$$(\beta_\mu\partial_\mu + M)\psi = 0, \quad (28)$$

where corresponding Kemmer-Duffin-Petiau matrices $\beta_\mu$, no matter what $5 \times 5$ or $10 \times 10$ matrices, are always independent of i.

If bosons and fermions exist together, this case corresponds to complex number a+bi. A unified complex number corresponds to super-unification of bosons and fermions. So Eq.(5) can be simplified to N= $(a+ib)^2 = a^2 - b^2 + 2abi$. The imaginary unit i may be represented as a form of matrix

$$\overline{\sigma}_2 = \begin{pmatrix} 0 & -1 \\ 1 & 0 \end{pmatrix}.$$

In Eq.(4) i or $\overline{\sigma}_2$ corresponds to interchange two types of particles. It corresponds to rotational angle of supersymmetry. $a+bi = \rho\exp(i\vartheta)$. $\rho = \sqrt{a^2+b^2}$ is modular of particle number, $\vartheta$ is an interchange factor. So the commutation relations are

$$\Phi_1\Phi_2 - N\Phi_2\Phi_1 = \Phi_1\Phi_2 - (a^2 - b^2 + 2abi)\Phi_2\Phi_1 = 0. \qquad (29)$$

When a=1 and b=0 is commutation relation; when a=0 and b=1 is anticommutation relation.

If real and imaginary units are unified by f=(1,i), Bose-Einstein (BE) distribution and Fermi-Dirac (FD) distribution are unified to

$$a_l = \frac{\varpi_l}{\exp(\alpha + \beta\xi_l) - f^2}. \qquad (30)$$

We discussed unified representation of two types of quantum statistics [39], i.e., supersymmetrical form. For a unified grand partition function is

$$\varsigma = -\frac{1}{f^2}\sum_l \omega_l \ln[1 - f^2 \exp(-\alpha - \beta\xi_l)]. \qquad (31)$$

Mohapatra discussed possible small violation of Pauli exclusion principle from parastatistics and infinite statistics, where bosons and fermions are unified [40]. The test of Pauli exclusion principle was proposed the first by Santilli in 1978, and then developed [41-43]. Based on the experiments and theories of particles at high energy, we proposed that particles at high energy possess a new unifying quantum statistics [39,36]. This super-unification connects naturally with supersymmetrical forms. Such the Pauli exclusion principle at high energy may not hold [39,40,44,45], which connects also with nonlinear quantum theories [36,46]. Further, non-Abelian gauge group not only is a nonlinear theory, and the equations of superfield are fundamentally nonlinear. Even supersymmetry may be related directly with nonlinear Born-Infeld-Higgs theory [47].

From 4-dimensional Wess-Zumino model in 1974 [48] to the superspace, usual space-time coordinates combining the parameterized spin variables construct 8-dimensional space. The string model is generally 26 dimensional spaces. The superstring combining Kaluza-Klein theory is 10 or 10+1 dimensional space. They point out that various usual supersymmetry theories are in higher dimensional space [6,13,24]. S.Weinberg discussed supersymmetry algebras in higher dimensions in Chapter 32 of <The Quantum Theory of Field> [1]. We researched the fractal dimensional matrix and linear algebra, and corresponding mathematics and physics, which may be developed to fractal and the complex dimension extended from fractal [49]. From this the fractal relativity is discussed, which connects with self-similarity Universe and the extensive quantum theory. The space dimension has been extended from real number to superreal and complex number. Combining the quaternion, etc., the high dimensional time $ict \to ic_1t_1 + jc_2t_2 + kc_3t_3$ is introduced. Such the vector and irreversibility of time are derived. Then the fractal dimensional time is obtained, and space and time possess completely symmetry. It may be constructed preliminarily that the higher dimensional, fractal, complex and supercomplex space-time theory covers all. While Regge theory and Veneziano model are analytic extension to complex space. We proposed that the simplest gauge group of four-interactions unified is GL(6,C) [36], which is also in a complex 6-dimensional space. In this paper, we study problem by a method where the higher dimensional supersymmetrical field combines complex number. Probably, new statistics and new developed direction must be in higher dimensional complex space.